\documentclass[aps,prb,a4paper,floatfix,showkeys,reprint,superscriptaddress]{revtex4-1}
\usepackage{amssymb,amsmath,mathrsfs}
\usepackage{natmove}
\usepackage{natbib}
\usepackage[pdftex]{graphicx}
\usepackage{bm}
\usepackage{isomath}
\usepackage{siunitx}
\usepackage[bookmarks, 
plainpages=false]{hyperref}
\hypersetup{ 
    colorlinks=true,%
    citecolor=blue
}
\usepackage{graphicx}
\usepackage{cleveref}
\usepackage[nolist,nohyperlinks]{acronym}
\usepackage[normalem]{ulem}

\crefname{figure}{Fig.}{Figs.}
\newcommand{\Gneq}{\tensor{\mathcal{G}}^{\neq}}

\newcommand{\kpar}{k_{\parallel}}
\newcommand{\sub}[1]{_{\mathrm{#1}}}
\newcommand{\unit}[1]{\,\,{\si{#1}}}

\newcommand{\alphat}{\tensor{\alpha}}
\newcommand{\unitvec}[1]{\hat{\bm{#1}}}
\DeclareMathAlphabet{\mathpzc}{OT1}{pzc}{m}{it}

\begin{document}
\def \figpath {./}
\title{Dispersion of guided modes in two-dimensional split ring lattices}
\author{Per Lunnemann}
\affiliation{DTU Fotonik, Department of Photonics Engineering, \O stedsplads 343, DK-2800, Denmark}
\author{A. Femius Koenderink}
\email{fkoenderink@amolf.nl}
\affiliation{Center for Nanophotonics, FOM Institute AMOLF,
Science Park 104, 1098 XG Amsterdam, The Netherlands}
\date{\today}

\begin{abstract}
We present a semi-analytical point-dipole method that uses Ewald lattice summation to find the dispersion relation of guided plasmonic and bi-anisotropic modes in metasurfaces composed of 2D periodic lattices of arbitrarily strongly scattering magneto-electric dipole scatterers. This method takes into account all retarded electrodynamic interactions as well as radiation damping selfconsistently. As illustration we analyze the dispersion of plasmon nanorod lattices, and of  2D split ring resonator lattices. Plasmon nanorod lattices support transverse and longitudinal in-plane electric modes. Scatterers that have an in-plane electric and out-of-plane magnetic polarizability, but without intrinsic magnetoelectric coupling, result in two bands that are mixtures of the bands of electric-only and magnetic-only lattices. Thereby bi-anisotropy through mutual coupling, in absence of building-block bi-anisotropy, is evident. Once strong bi-anisotropy is included in each building block, the Bloch modes become even more strongly magnetoelectric. Our results are important to understand spatial dispersion and bi-anisotropy of metasurface and metamaterial designs.
\end{abstract}
\date\today

\maketitle

\section{Introduction}
Periodic structures of scatterers have a long standing history in photonics, traditionally in guise of diffraction gratings\cite{Wood1935,Fano1941,	Loewen1997} and photonic crystals\cite{Yablonovitch1987a,Joannopoulos1997}, and more recently in the context of plasmonics, metamaterials, and metasurfaces \cite{Linden2001,Zheludev2012a,Kildishev2013}.  In plasmonics with noble metal particles  that support localized resonances, periodic chains of particles with subdiffraction pitch were already proposed in 1998\cite{Quinten1998} and demonstrated in 2003\cite{Maier2003} as potential candidates for guiding signals in a deep subwavelength fashion through near-field dipole-dipole interaction.\cite{Quinten1998,Koenderink2006,Brongersma2000,Weber2004,Maier2003,DeWaele2007,Yang2008}  While transport in these systems is very lossy, the exact formalism to describe the guiding mechanism in presence of long range retarded dipole-dipole interactions has remained a topic of ongoing work \cite{Quinten1998,Brongersma2000,Weber2004,Koenderink2006,Fructos2011}. This is in part due to the associated mathematical intricacies\cite{Linton2010} and in part to the fact that transport along chains of Lorentzian dipole resonators transcends plasmonics in relevance.  Two-dimensionally periodic systems of plasmon particles have traditionally been studied in case of diffractive lattices, in which case grating anomalies hybridize with localized surface plasmon resonances to give very sharp spectral features \cite{Vecchi2009a,Vecchi2009,Auguie2008,Kravets2008}.  These features have been pursued for field-enhanced spectroscopies\cite{Aroca2013,Langguth2014}, sensing\cite{Anker2008,Acimovic2009}, as well as  improved solid-state light sources \cite{Rodriguez2012,Lozano2013,Zheludev2008}.   Since the advent of 2D metamaterial arrays the response of subdiffraction pitch lattices of resonant scatterers has gained significantly in relevance \cite{Kildishev2013,Polman2008,Zheludev2012a}. 

While experimental studies of 2D metamaterials and metasurfaces usually probe transmission and reflection for some definite incident polarization and wave vector,\cite{Lunnemann2013,Sersic2009,Enkrich2005} the  fundamental underlying property of a lattice must be its dispersion relation or band structure, which summarizes the existence of guided as well as leaky modes. The spectrum of leaky modes supported by a lattice of split rings, for instance, would explain  the origin of angle-dependent transmission and reflection features\cite{Rockstuhl2006}, and would form an excellent basis to understand spatial dispersion in attributed effective material constants\cite{SimovskiPRB2007,MenzelPRB2008,AluPRB2011}.  
Complementary to the leaky modes, the guided mode structure would also be relevant, for instance for the proposed `lasing spaser' \cite{Zheludev2008} where a 2D metamaterial lattice is immersed in a gain medium, or when coupling a localized fluorescent source to a lattice in the near field. In this case, the modes subject to most gain, or the modes with strongest coupling to the source, need not correspond to resonances identified in normal incidence scattering experiments.  Rather, any guided modes supported by the lattice could be excited in any experiment that does not a priori restrict or impose parallel wave vector. An understanding of the band structure of 2D lattices of magnetic and electric resonant scatterers is therefore important for metasurface research.  

Previously, modal band structures of a 2D lattice of electric dipolar spherical scatterers have been theoretically treated\cite{Fructos2011,Zhen2008}, while only the leaky modes of SRR lattices have so far been assessed through transmission calculations and compared to experiments\cite{Rockstuhl2006,Lunnemann2013,Sersic2009}
In this paper we present a method to calculate band structures for arbitrary lattices of arbitrary magneto-electric dipolar scatterers, and illustrate its properties for simple lattices of plasmon rods, as well as idealized split rings.

\section{Lattice response }\label{sec:theory}
\begin{figure}
\centering
\includegraphics[width=1\columnwidth]{\figpath 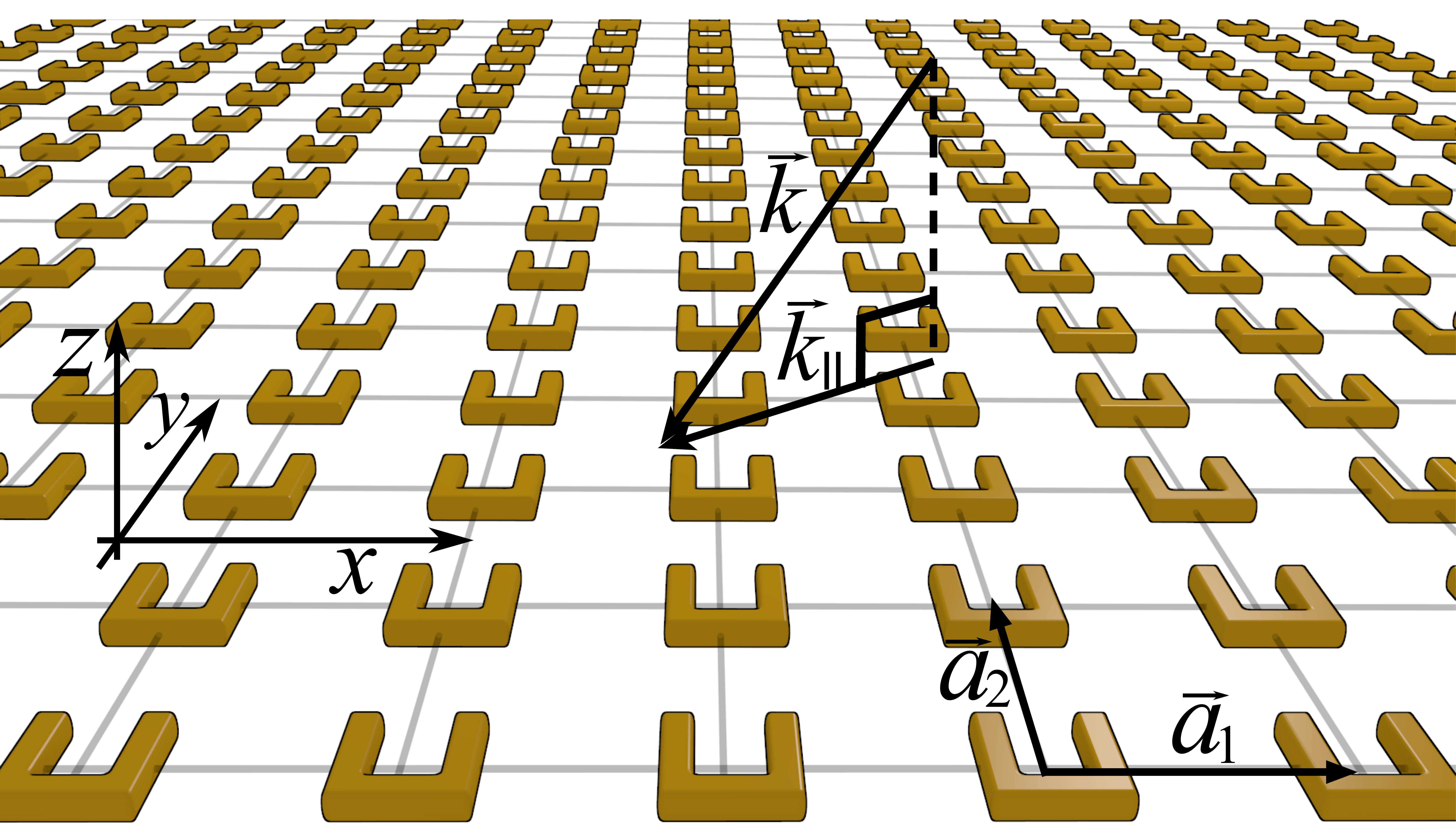}%
\caption{Illustration of the considered lattice, here sketched for split ring resonators, with a plane wave incident with an in-plane wave vector $k_{||}$.  \label{fig:setup}}%
\end{figure}

We consider a 2D lattice consisting of arbitrary magnetoelectric point scatterers in the dipole approximation, however, without making any electrostatic approximation. Each particle is described by a polarizability tensor, $\tensor{\alpha}$ that relates the induced electric and magnetic dipole moment, $\bm{p}$ and $\bm{m}$, to a driving electric and magnetic field
$\bm{E}$ and $\bm{H}$
according to\cite{Lindell1994,Sersic2011}
\begin{equation}
\begin{pmatrix}
  \bm{p} \\
  \bm{m}  \\
\end{pmatrix}=\alphat \begin{pmatrix}
  \bm{E} \\
  \bm{H}  \\
\end{pmatrix}.
\end{equation}
The magnetoelectric polarizability may be decomposed as
\begin{equation}
\alphat=
\begin{pmatrix}
  \alphat_{EE} & \alphat_{EH} \\
  \alphat_{HE} & \alphat_{HH}
\end{pmatrix},
\end{equation}
where $\alphat_{EE}$ is the $3\times 3$ electric polarizability tensor that quantifies the induced electric dipole moment in response to an electric field. Similarly, $\alphat_{HH}$ describes the magnetic polarizability that quantifies the induced magnetic dipole in response to a magnetic driving field. Finally, $\alphat_{EH}=-\alphat_{HE}^T$ denotes the magnetoelectric coupling that describes the induced electric dipole moment in response to a magnetic field and vice versa. This element controls the bi-anisotropy\cite{Lindell1994} of the medium giving rise to  chiral extinction under oblique incidence\cite{Sersic2011,Sersic2012}. We shall denote $\alphat$ the bare polarizability, since it describes the induced dipole moments in the absence of neighbouring point scatterers.
$\alphat$ is subject to  reciprocity and energy conservation constraints discussed in Ref. \onlinecite{Belov2003,Sersic2011}. We construct the electrodynamically consistent polarizability of a single scatterer, bound by the optical theorem, by addition of radiation damping
\begin{equation}
\alphat^{-1}=\alphat_{0}^{-1}
-\frac{2}{3}k^3 i \mathbb{I},\label{eq:dynPol}
\end{equation}
to an  electrostatic bare polarizability tensor $\alphat_{0}$ which can for instance be derived from an LC model. Here $k$ denotes the wave number, $\mathbb{I}$ is the 6-dimensional identity tensor and $.^{-1}$ denotes  matrix inversion.  In this work we will illustrate band structure calculations by considering  a specific group of scatterers representative of plasmon rods and of many metamaterial scatterers like \acp{SRR}.  In particular, we assume the only available responses to be electric along the $x$-direction (along the bar, or split, see fig. \ref{fig:setup}.) and/or magnetic along the $z$-direction (direction through the SRR loop), setting all other tensor elements to zero. I.e.
\begin{equation}
\tensor{\alpha}_{0}=\mathcal{L}(\omega)
\begin{pmatrix}
\eta_{E} & 0& \ldots & 0 & i\eta_{C}\\
0 & 0 & & & 0\\
\vdots & &\ddots & & \vdots\\
0 & & & 0 & 0\\
-i\eta_{C} & 0 & \ldots &0 & \eta_{H}
\end{pmatrix},\label{eq:barePol}
\end{equation}
where $\mathcal{L}(\omega)$ is a Lorentzian pre-factor 
\begin{equation}
\mathcal{L}(\omega)=V\frac{\omega_{0}^{2}}{\omega_{0}^{2}-\omega^{2}-i\omega\gamma},
\end{equation}
typical for a plasmon resonance or LC circuit model, where $V$ is the physical volume of the scatterer, $\gamma$ is the damping rate due to Ohmic losses,  $\omega_{0}$ denotes the resonance frequency, and $\eta_{E,H,C}$  are real dimensionless parameters that for LC circuits can be calculated from geometry.
 Recalling that the extinction cross section of a simple scatterer with  scalar polarizability $\alpha$ is $\sigma\sub{ext}=4\pi k \mathrm{Im}(\alpha)$, we note that for  tensorial polarizability, the extinction cross section varies with incidence condition, but is always  a linear combination of the imaginary part of the tensor eigenvalues, $\alpha_1$ and $\alpha_2$, of eq. \eqref{eq:dynPol}.
 The corresponding eigenvectors may possess a magneto-electric character, having both a component along $p_{x}$ and $m_{z}$ through intrinsic magneto-electric coupling when $\eta_{C}\neq0$ \cite{Sersic2011}.  Eigenvectors with  a phase-offset between $p_{x}$ and $m_{z}$ results in a `pseudochirality', \textit{i.e.}, a handedness-dependent extinction for some incidence angles.  We note that the \emph{maximum} value $\eta_C$ can attain is $\sqrt{\eta_E\eta_H}$, at which point one of the two eigen-polarizabilities reaches $0$ and the other reaches ${\cal L}(\omega) [\eta_E+\eta_H]$. For $LC$-circuit scatterers, maximum cross coupling is the norm, while removing cross coupling is a challenge.

Based upon the discrete dipole approximation method (DDA){\cite{Draine1994}}, the optical response of 2D periodic lattices of electric polarizabilities was didactically reviewed by de Abajo\cite{GarciadeAbajo2007} and extended to the full magneto-electric case in Refs. \onlinecite{Lunnemann2013, Kwadrin2014}. For consistency we  recapitulate the main findings. Consider  a 2D periodic lattice of point scatterers placed at $\bm{R}_{mn}=m\bm{a}_1+n\bm{a}_2$ (where $m$ and $n$ are integers, and $\bm{a}_{1,2}$ are the real space basis vectors, see \cref{fig:setup}).
The response of a particle at position $\bm{R}_{mn}$ is self-consistently set by the
incident field, plus the field of all other dipoles in the lattice according to\cite{GarciadeAbajo2007}
\begin{eqnarray}
\begin{pmatrix}
  \bm{p}_{mn} \\
  \bm{m}_{mn}  \\
\end{pmatrix}
&=&\alphat \left[
\begin{pmatrix}
  \bm{E}_{\mathrm{in}}(\bm{R}_{mn}) \\
  \bm{H}_{\mathrm{in}}(\bm{R}_{mn})  \\
\end{pmatrix} \right. \nonumber \\ & & \quad \left. + \sum_{m' \neq m, n'\neq n}
\tensor{G}^0(\bm{R}_{mn}-\bm{R}_{m'n'})
\begin{pmatrix}
  \bm{p}_{m'n'} \\
  \bm{m}_{m'n'}  \\
\end{pmatrix}
 \right] \nonumber \\
\end{eqnarray}
where $\tensor{G}^0(\bm{R}_{mn}-\bm{R}_{m'n'})$  is the $6\times 6$ dyadic Green function of the medium surrounding the lattice. 
Plane wave incidence with parallel wave vector $\bm{k}_{||}$ allows a Bloch wave form $(\bm{p}_{mn},\bm{m}_{mn})^T
  =e^{i\bm{k}_{||}\cdot\bm{R}_{mn}}(\bm{p}_{00},
  \bm{m}_{00})^T$ to obtain
\begin{equation}
\begin{pmatrix}
  \bm{p}_{00} \\
  \bm{m}_{00}  \\
\end{pmatrix}
=[\alphat^{-1} -
\Gneq(\bm{k}_{||},0)]^{-1}
\begin{pmatrix} \bm{E}_{\mathrm{in}}(\bm{R}_{00}) \\
  \bm{H}_{\mathrm{in}}(\bm{R}_{00})  \\
  \end{pmatrix}
  \label{eq:effpol}
\end{equation}
Here, $\Gneq(\bm{k}_{||},0)$ is a summation of the
dyadic Green function $\tensor{G}^0$ over
all positions in the lattice barring the
origin:
\begin{equation}
\Gneq(\bm{k}_{||},\bm{r})=\sum_{m\neq 0,n\neq 0}
\tensor{G}^0({\bm{R}}_{mn}-\bm{r})e^{i\bm{k}_{||}
\cdot \bm{R}_{mn}}\label{eq:latticeSum}
\end{equation}
In this work, we take the surrounding medium, that defines $\tensor{G}^0$,  to be homogeneous. Implementation of the sum of $\tensor{G}^0$ was carried out using the Ewald lattice summation technique\cite{Linton2010}, that consists of splitting a poorly convergent sum, like eq. \eqref{eq:latticeSum}, into two exponentially convergent sums as summarized in Refs. \onlinecite{Lunnemann2013,Kwadrin2014}. The same techniques, can be extended to lattices in stratified dielectric systems, complex unit cells,  and stacks of lattices\cite{Kwadrin2014}.

The factor $[\alphat^{-1} -\Gneq(\bm{k}_{||},0)]^{-1}$ in Eq.~(\ref{eq:effpol}) is  identified as an \emph{effective} polarizability tensor, $\tensor{\alpha}\sub{eff}$, of the scatterer, renormalized by the scattering from all other lattice sites. In a lossless system the Bloch wave dispersion would correspond to those frequencies for which 
$\det(\tensor{\alpha}^{-1}-\Gneq)=0$,
or equivalently, those frequencies for which $\alpha\sub{eff}$ has a pole. These two dimensional lattices in fact have radiative (if $k_{||}\leq\omega/c$) and Ohmic loss. For lossy systems a real dispersion relation is not defined, and one should in principle seek either complex wave vector - real frequency solutions, or conversely complex frequency - real wave vector solutions for which $\det(\tensor{\alpha}^{-1}-\Gneq)=0$  as first noted by Barker and Loudon.\cite{Loudon1972,Koenderink2006} We simply evaluate $\tensor{\alpha}^{-1}-\Gneq$ for real $\omega$ and real $\kpar$. In particular we consider the imaginary part of the  eigenvalues of $[\tensor{\alpha}^{-1}-\Gneq ]^{-1}$ as they directly relate to extinction.  Distinct bands emerge, the width of which we identify as the damping rate \cite{Zhen2008} due to  both Ohmic and radiation damping.
For the effectively $2\times 2$ form of the single-particle polarizability that we use in this paper (Eq. \eqref{eq:barePol}), for each $(\omega,\kpar)$-point we obtain at most two nontrivial eigenvectors and eigenvalues.
To separate the dispersion in bands, we group eigenvalues by continuity of the projection of the corresponding eigenvectors with the eigenvectors of neighboring $(\omega,\kpar)$-points. 

\section{Results}
As illustration we consider four types of scatterers, starting with plasmon rods and culminating at a realistic description for split rings.  It has been experimentally demonstrated that SRRs are well described with a dipolar polarizability as in eqs. \eqref{eq:barePol} and \eqref{eq:dynPol} provided one takes bi-anisotropy $\eta_{C}$ at the upper limit $\sqrt{\eta_{E}\eta_{H}}$ \cite{Lunnemann2013,Sersic2012,Sersic2011}.
To help understand what effect magnetoelectric coupling has,  we consider three sub-cases prior to analyzing the SRR lattice with maximum intrinsic coupling. These consist of (1) plasmon rods along $x$  (only $\eta_E\neq0$),  (2) magnetic dipolar antennas along $z$ (only $\eta_H\neq0$) and (3) uncoupled SRRs without bi-anisotropy (setting $\eta_E=\eta_H=1$ and $\eta_C=0$).  Throughout the paper we use parameter values as stated in table \ref{tab:parameters}.
\begin{table}
\caption{\label{tab:parameters} Chosen parameter values.}
\begin{ruledtabular}
\begin{tabular}{lll}
  \textbf{Parameter} & \textbf{Value} & Description\\\hline
  $V$ & $(80\unit{nm})^{3}$ &Physical volume of scatterer.\\
$\gamma$ & $1\cdot10^{12}\unit{s^{-1}}$ & Damping rate.\\
$a_{1}=a_{2}$ & $300 \unit{nm}$ & Lattice constant.\\
$\omega_{0}$ & $2\pi c/1.5\unit{\micro\meter}$ & Resonance frequency
\end{tabular}
\end{ruledtabular}
\end{table}
To more clearly resolve the modes, we take an Ohmic damping rate $\gamma$  $\sim$ 100 times less than that of gold\cite{Johnson1972}. The  parameters yield an extinction cross section per scatterer of $0.2-0.3\unit{\micro\meter^2}$, comparable to measured values\cite{Husnik2008,Lunnemann2013}.

\subsection{Scalar anisotropic scatterers}\label{sec:scalarScat}
\begin{figure}
\centering
\includegraphics[width=1\columnwidth]{\figpath 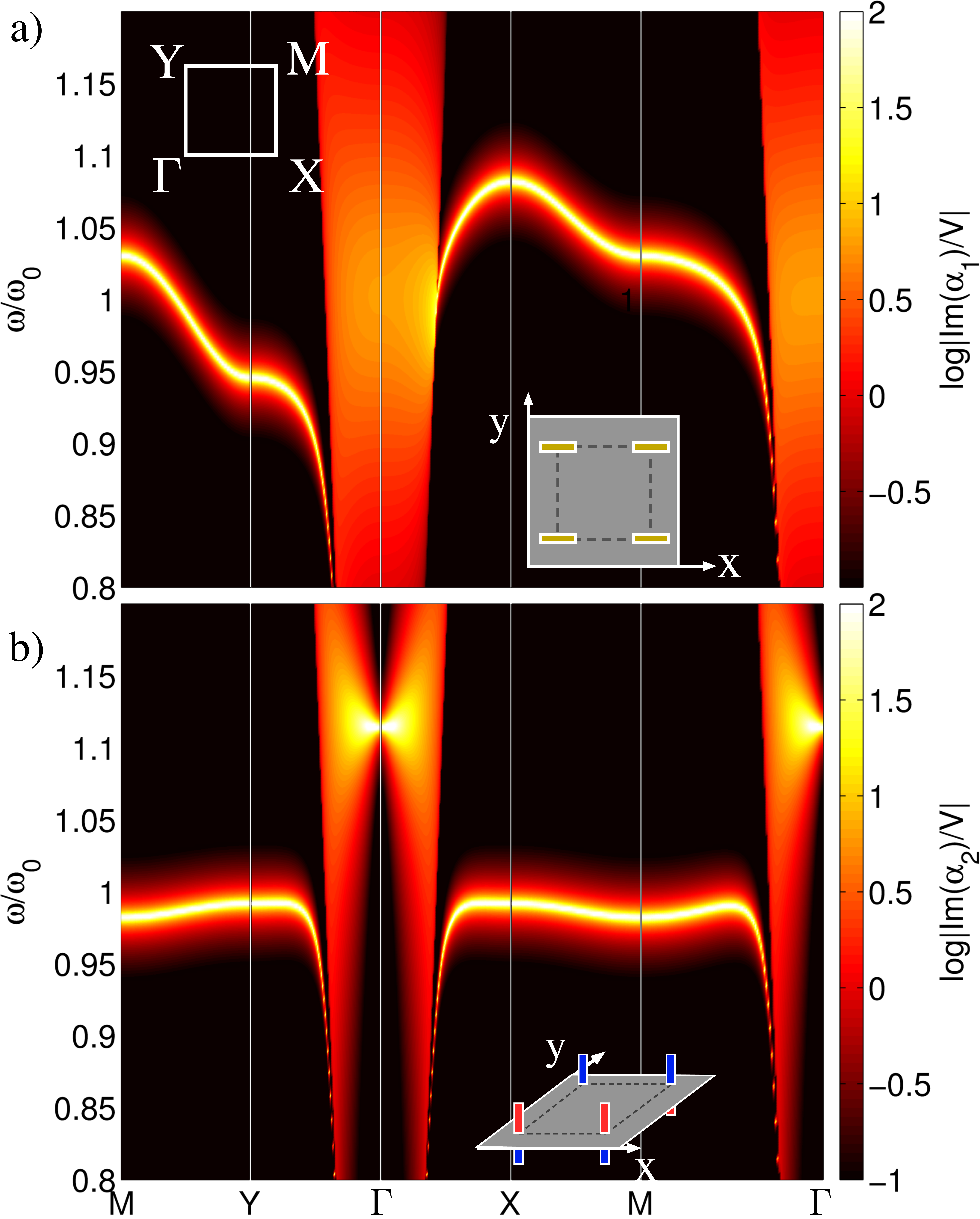}%
\caption{(Color online)
Calculated effective polarizability, $\log[\mathrm{Im}(\alpha)/V]$ as a function of $k_{||}$ and $\omega$ for a) an in-plane bar-type electric structure as illustrated in the inset with $\eta_H=\eta_C=0$ and $\eta_E=1$ in eq. \eqref{eq:barePol} and b) a transverse bar-type magnetic structure as illustrated in the inset with with $\eta_E=\eta_C=0$ and $\eta_H=1$ in eq. \eqref{eq:barePol}. \label{fig:dispBarX}}%
\end{figure}

We first consider plasmon rods with bare polarizability given by eq. \eqref{eq:barePol} and setting $\eta_H=\eta_C=0$ and $\eta_E=1$.
In figure \ref{fig:dispBarX}a) we present the calculated dispersion diagram sweeping $(k_x,k_y)$  through  the reduced Brillouin zone along the following path:  $(\pi/d,\pi/d)\rightarrow (0,\pi/d)\rightarrow (0,0)\rightarrow (\pi/d,0)\rightarrow (\pi/d,\pi/d)\rightarrow (0,0)$ also denoted $\mathrm{M}\rightarrow\mathrm{Y}\rightarrow\mathrm{\Gamma}\rightarrow\mathrm{X}\rightarrow\mathrm{M}\rightarrow\mathrm{\Gamma}$.  
The discontinuity  on each side of $\mathrm{\Gamma}$ indicates the light line. The sharp resonances below the light line are guided modes of the lattice. Two distinct modes are observed on either side of the $\mathrm{\Gamma}$-point. Within the domain $\mathrm{\Gamma}\rightarrow\mathrm{Y}$, the dipole phase is constant along $\hat{\bm{x}}$ and varies along $\hat{\bm{y}}$, transverse to the dipole moment orientation. Thus, this mode is a \ac{TIE} mode, with dipole moments perpendicular to $\kpar$. The heads to tail arrangement along $\hat{\bm{x}}$ and alternating direction when going along $\hat{\bm{y}}$ results in a redshift of the resonance as has previously been explained using simple hybridization models for electrostatic 1D and 2D systems \cite{Brongersma2000,Koenderink2006,Weber2004,Prodan2003,Liu2009}. In such a hybridization model, resonance shifts of coupled dipoles can be understood from considering the quasi-static interaction energy, $U$, between two dipoles $\bm{p}_1$ and $\bm{p}_1$ separated by a vector $\bm{r}$ 
\begin{equation}
U\propto\frac{\bm{p}_1\cdot\bm{p}_2-3[\bm{p}_1\cdot\hat{\bm{r}}][\bm{p}_2\cdot\hat{\bm{r}}]}{r^3}.\label{eq:dipoleEn}
\end{equation}
Accordingly,  longitudinal coupling of parallel (antiparallel) dipoles leads to redshifts (blueshifts),  while for transversely coupled dipoles the coupling strength is reduced and opposite in sign. As caveat we note that such hybridization models strictly apply only in electrostatics, whereas here we treat retarded interactions between lossy, resonant dipoles.

Analogously, within the domain $\mathrm{\Gamma}\rightarrow\mathrm{X}$ all dipoles point along the wave vector, and the mode is therefore \ac{LIE}\cite{Zhen2008}. In this configuration, the head to head arrangement along $\hat{\bm{x}}$ and the fixed direction along $\hat{\bm{y}}$ results in a blueshift. For the regions $\mathrm{Y}\rightarrow \mathrm{M}$ and $\mathrm{X}\rightarrow \mathrm{M}$ the mode possesses a mixed  transverse and longitudinal character. 
For wave vectors within the light cone, we notice a faint resonance. This resonance is very broad due to radiation damping, and is the resonance that is probed in farfield transmission spectra \cite{Sersic2009,Lunnemann2013,Enkrich2005}. Compared to the single particle radiative linewidth, the collective resonance linewidth is broader by more than an order of magnitude. This collective superradiant damping effect has been observed experimentally in density dependent studies of  transmission at normal incidence for 2D lattices.\cite{Sersic2009} Since the time-averaged far field flux from a single dipole pointing along $\unitvec{x}$ is proportional to $\sin^{2}(\theta)$ with $\theta$ being the azimuthal angle along $x$ \cite{Bohren1983}, scattering out of the lattice plane is strong for in-plane modes. For longitudinal modes this damping monotonically reduces with increasing $k_{||}$ as dipoles do not radiate along their axis \cite{Weber2004,Koenderink2006}. For transverse modes the radiation damping is constant or increases with $k_{||}$ when approaching the light line. 
Finally, the modes far below the light line have a constant width comparable to the single particle Ohmic damping rate $\gamma$. However, very close to the light line, the damping rates drop well below the Ohmic damping rate indicating that the modes near the light line are very weakly confined and have almost no mode overlap with the metal scatterers. As is the case for, e.g.,  a thin dielectric slab in a symmetric host environment,  even a plane of weakly polarizable particles binds a guided mode, however with a very large fraction of its energy density in air. We note the strong similarities to the calculated dispersion of 1D as well as 2D lattices of spherical electric scatterers\cite{Weber2004,Koenderink2006,Zhen2008}.

We now turn to the  case of out-of-plane magnetic scatterers with  bare polarizability set by $\eta_E=\eta_C=0$ and $\eta_H=1$ in eq. \eqref{eq:barePol}. This is a hypothetical case as magnetic scatterers are not readily available.  Yet  very high index dielectric spheres and spheroids have magnetic dipole resonances, so that realizations could be envisioned~\cite{Krasnok}.
In fig. \ref{fig:dispBarX}b) we present the calculated effective polarizability. The dispersion is symmetric about $\mathrm{\Gamma}$ owing to the rotational symmetry of this lattice. Obviously, the observed mode is a \ac{TM} mode. As opposed to the \ac{TIE} and \ac{LIE} modes in fig. \ref{fig:dispBarX}, we can clearly resolve a resonance above the light line. 
For a single transverse dipole the radiated intensity perpendicular to the lattice is zero, while radiation in the plane is strong \cite{Bohren1983}. Hence for the array at $\kpar =0$ there is no radiative loss. 
The increasing broadening  when going from $k_{||}=0$ to the light line, was also claimed for 1D chains of plasmonic particles\cite{Weber2004,Koenderink2006}. As the wave vector sweeps to the light line, the radiative loss, which by momentum conservation has the same in-plane wave vector, has an increasingly good overlap with the single dipole radiation pattern, thereby causing the radiative loss to increase from $0$ at $k_{||}=0$ to large values.

Comparing fig. \ref{fig:dispBarX}a) and \ref{fig:dispBarX}b), the \ac{TIE} and \ac{TM} modes are seen to converge asymptotically to the light line, while the \ac{LIE}-mode crosses the light line. For the transverse modes all dipoles are perpendicular to the propagation vector and therefore couple strongly to free photons with an  anti-crossing as a result. For the \ac{LIE}  mode all dipoles are parallel to the wave vector and thus hindered from coupling to the far field. Consequently no anti-crossing is observed in good agreement with previous results on 1D and 2D arrays of scatterers\cite{Koenderink2006,Weber2004,Zhen2008}. Finally, we note that the modes of the out-of-plane magnetic antennas are much less dispersive than those of the electric in-plane antennas showing an almost flat band between Y-M and X-M. Inspecting the interaction energy in the magnetostatic equivalent of eq. \eqref{eq:dipoleEn}, we note that for the TM mode near Y (X) a blueshift is induced from adjacent parallel dipoles along $\hat{\bm{x}}$  ($\hat{\bm{y}}$) while a redshift of equal magnitude is induced from adjacent anti-parallel dipoles along $\hat{\bm{y}}$ ($\hat{\bm{x}}$), leading to net cancellation of the hybridization energies. This cancellation of nearest-neighbor contributions  holds for any periodicity, i.e., also for retarded interactions.  To the contrary,  for the electric in-plane antennas near Y (X), the heads to tail arrangement along $\hat{\bm{x}}$ ($\hat{\bm{y}}$) and anti-parallel arrangement along $\hat{\bm{y}}$ ($\hat{\bm{x}}$) both contribute with a red  (blue) shift. Hence any path connecting Y and X exhibits a larger variation in frequencies compared to the TM mode.

\subsection{Split ring resonators}
We now turn our attention to \ac{SRR}-type scatterers.
We describe the per-building block magnetoelectric coupling by the parameter $\eta_{C}$ in eq. \eqref{eq:barePol}. Energy conservation dictates that $\eta_{C}$ is bound by  $|\eta_{C}|\leq\sqrt{\eta_{E}\eta_{H}}$,\cite{Sersic2011} where  equality holds for a truly planar scatterer described as a single resonant circuit. It has been  demonstrated experimentally, and by full-wave simulations\cite{Lunnemann2013,Sersic2012,Sersic2011,Arango2013} that real SRRs indeed possess a coupling strength $\eta_{C}$ close to the upper limit $\sqrt{\eta_{E}\eta_{H}}$. We note, that for more complicated scatterers though, i.e. nested split rings,  lower cross couplings can occur \cite{Sersic2012,Arango2013}. For clarity we shall first consider the two cases of absent and partial coupling between the magnetic and electric dipole by setting $\eta_C=0$ and $\eta_C=0.5\sqrt{\eta_{E}\eta_{H}}$, respectively, in eq.\eqref{eq:barePol} before finally considering the realistic case with full cross coupling $\eta_C=\sqrt{\eta_{E}\eta_{H}}$.
The calculated dispersion diagrams are presented in figure \ref{fig:SRRCouplingDep}.

\subsubsection{SRR without cross coupling (anisotropic)}\label{sec:SRRnoCoup}
\begin{figure*}
\centering
\includegraphics[width=1\textwidth]{\figpath 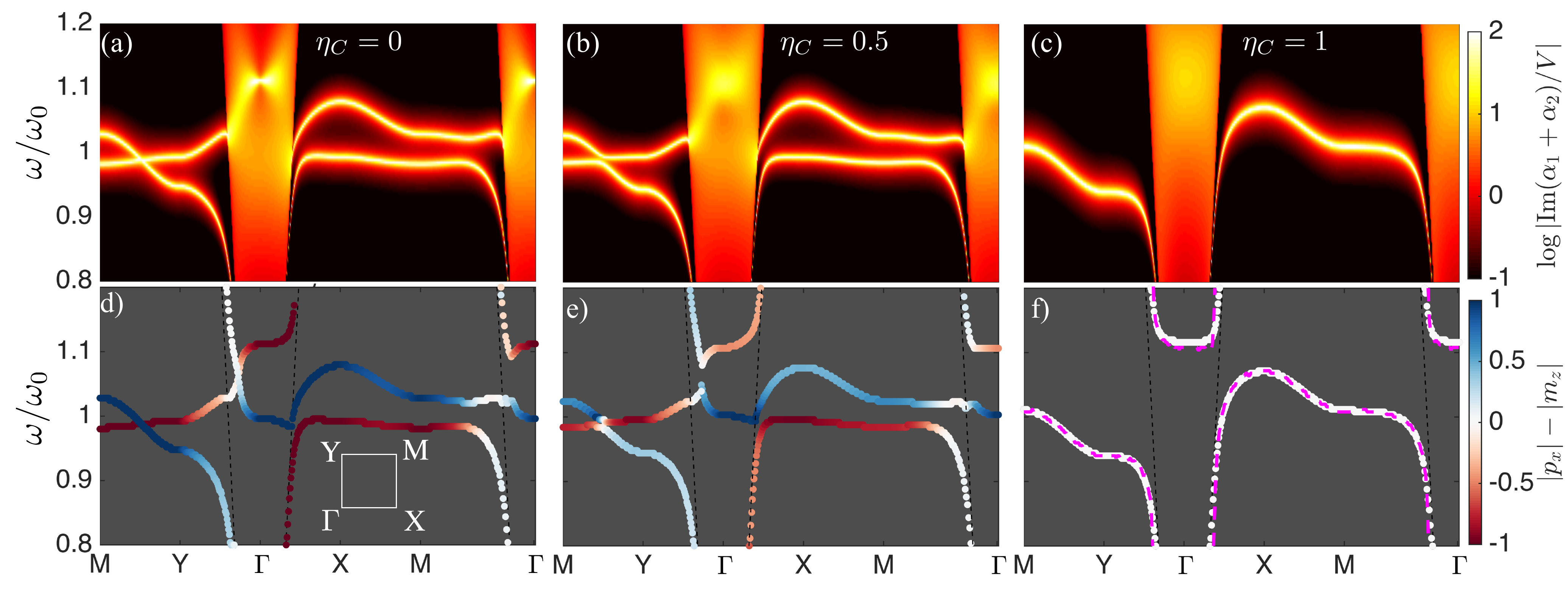}%
\caption{(Color online) a-c) Calculated sum of eigenvalues, $\log[\mathrm{Im}(\alpha_{1}+\alpha_{2})/V]$, as a function of wave vector, $k_{||}$, and normalized frequency $\omega/\omega_0$ for a SRR lattice with a) cross coupling $\eta_{C}=0$ and b) $\eta_{C}=0.5$, and c) $\eta_C=1$. d-f) calculated mixing ratio $ \zeta=|p_x|-|m_z|$ of the eigenvectors for the extracted bands. Blue corresponds to a pure electric dipole while red corresponds to  pure magnetic dipole. White is a balanced mix of magnetic and electric dipoles. Dashed black lines indicate the light lines. Magenta dashed line in f) indicates the sum of frequency shifts $\Delta\omega_E(\kpar)+\Delta\omega_H(\kpar)+\omega_0$ of the two modes in d). \label{fig:SRRCouplingDep}}%
\end{figure*}

We start by considering the case of no electro-magnetic cross coupling, $\eta_{C}=0$ and $\eta_{E}=\eta_{H}=1$ . In this case, for each $(\kpar,\omega)$-pair two non-trivial eigenvalues exist. In fig. \ref{fig:SRRCouplingDep}a) the calculated sum of imaginary part of eigenvalues, $\mathrm{Im}\alpha_1+\mathrm{Im}\alpha_2$ is plotted.
We immediately identify that the dispersion diagram resembles the superposition of those in fig. \ref{fig:dispBarX} for the purely electric, and purely magnetic objects. One mode traces the in-plane mode in fig. \ref{fig:dispBarX}a) while we observe some differences  between the other mode and the  \ac{TM} mode in fig. \ref{fig:dispBarX}b), especially for the region M-Y-$\mathrm{\Gamma}$. These differences arise from inter-particle coupling between electric dipoles, magnetic dipoles and between electric and magnetic dipoles. 

To clarify how fig. \ref{fig:SRRCouplingDep}a) and fig. \ref{fig:dispBarX}b) (TM mode) differ  we define a electromagnetic mixing ratio as
\begin{equation}
\zeta_{j}=|p_{j,x}|-|m_{j,z}|,\label{eq:mixingR}
\end{equation}
where $p_{j,x}$ ($m_{j,z}$) is the electric $x$ (magnetic $z$) component of the $j$th normalized eigenvector, i.e. $\sqrt{|p_{j,x}|^2+|m_{j,z}|^2}=1$. For $\zeta=-1$ the mode is purely TM while for $\zeta=1$ the mode is purely in-plane electric. The mixing ratio of the bands in fig. \ref{fig:SRRCouplingDep}a) is presented in fig. \ref{fig:SRRCouplingDep}d). Generally the two modes are clearly identifiable as strongly electric resp. magnetic, as expected if the dispersion were that of an uncoupled electric lattice and magnetic lattice. In fact,
 the modes are \emph{purely} in-plane electric, respectively TM, at all the symmetry point M, Y, $\mathrm{\Gamma}$, and X. A strongly mixed character occurs close to the light line, in particular midway $\mathrm{\Gamma}$ and Y,  and at the light line crossing between $\mathrm{\Gamma}$ and M, commensurate with the fact that there the mode must match a plane wave propagating in the array plane, which  carries both out-of-plane $H$ and in plane $E$.

\begin{figure}
\centering
\includegraphics[width=1\columnwidth]{\figpath 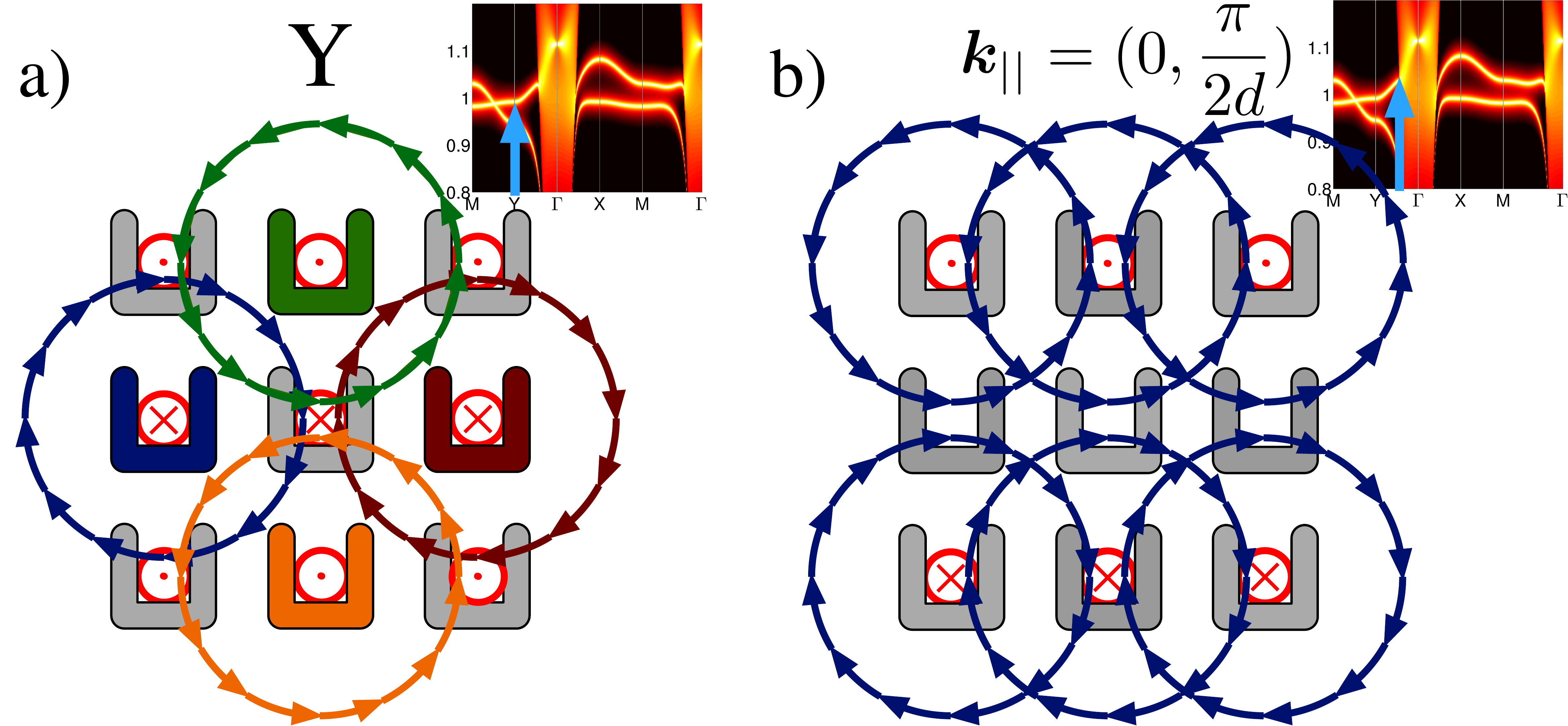}%
\caption{(Color online) Illustration of the distribution of magnetic dipoles and their associated electric fields in a SRR lattice with $\eta_C=0$ for wave vectors at a) $\mathrm{Y}$ and b) $\mathrm{Y/2}$ as indicated by a blue arrow in the dispersion map inset. Red circles with a dot (cross) indicate a magnetic dipole pointing out of (into) the paper.\label{fig:modeIllustration}}%
\end{figure}

Before we introduce cross coupling $\eta_C$, we consider the hybridization interaction between magnetic and electric dipoles. In fig. \ref{fig:modeIllustration} we illustrate the spatial distribution of magnetic dipoles and their associated electric fields at the point in $\kpar$-space $\mathrm{Y}=(0,\pi/2d)$ and 
the point midway between $\mathrm{\Gamma}$ and Y, denoted as $Y/2$.
Considering the field lines at Y, depicted in fig. \ref{fig:modeIllustration}a),  at the location of the central SRR the electric fields from adjacent magnetic moments cancel. We therefore conclude that at Y  the lattices of magnetic and electric dipoles are essentially decoupled,  in good agreement with the unit value of the mixing ratio in fig. \ref{fig:SRRCouplingDep}d). A similar analysis holds for the X-point. 
Considering the point Y/2 in fig. \ref{fig:modeIllustration}b), the electric field lines of magnetic dipoles adjacent to a central site add up along $\hat{\bm{x}}$ and from this we infer that the \ac{TIE} and \ac{TM} mode strongly mix at Y$/2$. This is in contrast to the point X/2, where the field lines add up along $\hat{\bm{y}}$, along which the SRR is not polarizable. Hence no dipole moment is induced and one mode is therefore purely TM, while the other is purely electric in nature. The same results for the absence/presence of magnetoelectric crosscoupling is obtained by starting the  hybridization analysis from magnetic fields due to  in-plane electric dipoles rather than vice versa, as expected from reciprocity.

To conclude, even if one starts with building blocks that have no magnetoelectric coupling, once placed in a dense lattice, the collective modes have mixed electric and magnetic character except at symmetry points.  

\subsubsection{SRR with cross coupling (bi-anisotropic)}
Most realized metamaterial lattices will consist of building blocks possessing an intrinsic magneto-electric coupling that couples the excitation of electric and magnetic dipoles in a single building block according to a definite amplitude and phase relation.
We  consider  partial electromagnetic cross coupling setting $\eta_C=0.5$ and $\eta_E=\eta_H=1$  in fig. \ref{fig:SRRCouplingDep}b). Comparing with the case $\eta_C=0$ in fig. \ref{fig:SRRCouplingDep}a), we immediately see a close resemblance, apart from a clearly resolved anti-crossing midway between M and Y. The calculated mixing ratio (see figure \ref{fig:SRRCouplingDep}e)) evidences that the modes are no longer purely electric or purely magnetic at any of the symmetry points. Near the anti-crossing , the mixing ratio becomes 0 implying that the two modes carry equal electric and magnetic content. 
The associated complex phase difference defined as
\begin{equation}
\Delta\phi_{j}=\arg{p_{j,x}}-\arg{m_{j,z}}
\end{equation}
is, at the anti-crossing point, found to be $\Delta\phi_{1}=\pi/2$ for the lower band and $\Delta\phi_{2}=-\pi/2$ for the upper band.
This distinct phase difference  between the two anticrossing bands  signifies that   one solution  has  the electric dipole a quarter cycle in advance of the magnetic one, while for the other solution the electric dipole lags the magnetic dipole by a quarter cycle. This distinction stems from the physics intrinsic to the single bi-anisotropic scatterer,  wherein  the polarizability tensor of  a cross-coupled scatterer has two distinct eigenvalues,  corresponding to high, respectively low scattering strength, with eigensolution corresponding to either an advanced or lagging electric dipole relative to the magnetic response.    
In $LC$-circuit terms, a strong difference in response to a driving field with either an advance or lag can be understood by noting that in the driving $E_x d  +  i\omega H_z A$ ($d$ the capacitor gap,  $A$ the loop area) the electric term driving the capacitor and the electromotive force due to a changing flux, either add or cancel depending on phase.   In a scattering experiment  this results in a strongly handed response under oblique incidence,  since oblique incidence circular polarization carries a phase difference between $E_x$ and $H_z$ \cite{Sersic2012,Lunnemann2013}.  
Returning to the physics of the lattice, given the Bloch wave vector $\kpar=(\mathrm{M}+\mathrm{Y})/2$   one can explicitly  calculate the fields exerted on a central dipole  by all  its neighbors.  We indeed find that when $p_x=im_z$ respectively $p_x=-im_z$,  the overlap  of the fields  a central dipole scatterer  receives from its  neighbors in the lattice is very different in strength (addition resp. cancellation in terms of $E_x d  +  i\omega H_z A$).   

Finally, we consider the  maximally coupled case, $\eta_C=\eta_E=\eta_H=1$, presented in figure \ref{fig:SRRCouplingDep}c,f). 
The most remarkable aspect  is that one of the two bands vanishes, leaving only one band.  That this must happen is easily understood by noting that at maximum cross coupling $\eta_C = \sqrt{\eta_E \eta_H}$,   one of the two eigenvalues of the single-object polarizability tensor vanishes.  In other words, a maximally crosscoupled SRR has just one mode of oscillation and not two,  in which furthermore the relative phase and amplitude of $p$ and $m$ are locked.
A didactic example is an $LC$ circuit, which has just one resonant mode of oscillation  where the same circulating charge  gives rise to both $p_x$ and $m_z$ in a fixed phase and amplitude ratio. This intuition for the $LC$ circuit is only reconcilable with a $2\times 2$ polarizability tensor if one eigenvalue vanishes. By extension, this also means that a lattice of maximally cross-coupled scatterers presents only one band, not two. 
A remarkable observation is that, beyond the light line, the band structure is similar to the primarily in-plane mode of the lattice with $\eta_C=0$, seen in fig. 3a) and 3d). However, this similarity only holds for the $(\omega,\kpar)$-relation but not for the associated eigenfunctions, since the eigenfunctions necessarily show equally strong $p$ and $m$ resulting in $\zeta=0$ at the maximum cross coupling condition,  whereas $0 < \zeta\leq 1$ in the electric-only case. 

The resemblance of bandstructures traces back to the fact that for the fully coupled system the bandstructure expressed as a shift $\Delta\omega(k_{||})=\omega(k_{||})-\omega_0$ is closely connected to the \emph{sum} of the band structures of the uncoupled  electric and magnetic band in fig 3d). In figure 3f) the sum of frequency shifts for the two bands of the uncoupled system $\Delta\omega_E(k_{||})$+ $\Delta\omega_H(k_{||})+\omega_0$ is overplotted with the fully coupled result.  Since the magnetic mode is almost flat, the result is that the fully coupled system closely resembles the purely electric system in band structure.
The mathematical reasoning behind this summation argument, traces back to   analysis of the matrix form of the bare  polarizability Eq. (\ref{eq:barePol}). As crosscoupling approaches the  maximally coupled case,  one eigenvalue vanishes while the second eigenvalue simplifies to the sum of the diagonal contributions. Similar reasoning extends to the full effective polarizability that includes the lattice summation. Thereby,  one dispersion band, corresponding to the vanishing eigenpolarizability, converges to $\Delta\omega =0$ and vanishes in strength, while the second band approaches the sum of the dispersions in the uncoupled systems.

\section{Conclusion}
To conclude, we  discussed a method based on magnetoelectric point-dipole interactions and Ewald lattice summation to approximate the dispersion relation of two dimensional lattices of bi-anisotropic scatterers,  accounting  for all retarded electrodynamic interparticle interactions. Our results show that simple square lattices of plasmon rods that are dense, i.e. of subdiffraction pitch,  support a mode structure characterized by weakly confined guided modes with a dispersion very close to the light line for frequencies to the red of the single scatterer resonance, and tightly confined guided modes at wave vectors well away from the light line.  These modes are dispersive in a manner similar to results obtained previously for 1D plasmon chains, with the added complication that modes can have a mixed transverse and longitudinal character.  Lattices of scatterers that have a intrinsically decoupled electric and magnetic polarizability in each element, will have a dispersion in which modes have a mixed magneto-electric character. Furthermore, we  reported how introduction of bi-anisotropy in each building block modifies the dispersion. For full cross coupling a single mode prevails with the electric and magnetic dipoles being interlocked with equal magnitude and a fixed $\pi/2$ phase.

As outlook,  while we  presented results for simple square lattices of split ring type resonators  with just an in plane electric moment and out of plane magnetic moment,  the method is easily generalized to deal with arbitrarily complex multi-element lattices of arbitrary magnetoelectric scatterers, provided that the dipole approximation is met.  Thus our method is important for many structures, including  metasurface designs with complex unit cells that comprise many elements.  Since the method is fast, it should thus be possible to screen many different lattice symmetries and arrangements in the unit cell for desirable properties, such as minimized bi-anisotropy and spatial dispersion. As regards  actual measurements of such dispersion relations, we note that measurements in the visible domain would likely be hampered by the much strong Ohmic damping than assumed in this work, and the drawback that the Brillouin zone extends to very large wavevectors,  rendering even near-field microscopy impractical.  These are exactly the drawbacks that have made it impossible to verify the projected dispersion of 1D plasmon chains beyond the light line\cite{DeWaele2007,Yang2008}.  However, in the radio frequency domain, the  dispersion relations should be more readily available. The RF domain offers as advantages that low-loss split rings can be made,  that near-field probes with $\lambda/100$ resolution are routine, and  that both phase and amplitude can be mapped so that $k$ is directly measured. Finally we believe that the calculated dispersion relations should be important for lasing spaser experiments, in which lattices are studied in presence of gain to achieve lasing.  The modes with the best tradeoff between loss and confinement should be found close to and just below the light line, rather than at the $\kpar=0$ point,  which was proposed as the lasing mode originally \cite{Zheludev2008}. Furthermore, from work on Yagi-Uda antennas in the optical domain it is well known that directional scattering, and directional emission of embedded emitters is strongly linked to the modes of 1D particle chains just below the light line \cite{Kosako2010,Koenderink2009,Arango2012,Curto2010a}.  Similarly, for 2D arrays our work points at design strategies for shaping directional emission.

\begin{acknowledgments}
We are grateful to Yuntian Chen and Lutz Langguth for insightful comments. This work is part of the research program of the “Stichting
voor Fundamenteel Onderzoek der Materie (FOM)”, which
is financially supported by the “Nederlandse Organisatie
voor Wetenschappelijk Onderzoek (NWO)”. AFK. gratefully acknowledges a NWO-VIDI grant for financial support. PL acknowledges support by the Carlsberg Foundation as well as the Danish Research Council for Independent Research (Grant No. FTP 11-116740).
\end{acknowledgments}

\begin{acronym}
\acro{FDTD}{finite difference time domain}
\acro{SRR}{split ring resonator}
\acro{TIE}{transverse in-plane electric}
\acro{LIE}{longitudinal in-plane electric}
\acro{TM}{transverse magnetic}
\end{acronym}

\bibliography{Papers-SRRDispersionPaper}

\end{document}